\documentclass[12pt]{article}
\usepackage[centertags]{amsmath}
\usepackage{amssymb,amsfonts}
\usepackage{latexsym}
\usepackage{epsfig}
\usepackage{graphicx,psfrag,subfigure}
\usepackage{rotating}
\newcommand{\be}{\begin{equation}} \newcommand{\ee}{\end{equation}}
\newcommand{\bea}{\begin{eqnarray}} \newcommand{\eea}{\end{eqnarray}}
\newcommand{\bse}{\begin{subequations}} \newcommand{\ese}{\end{subequations}}

\begin{document}
\begin{center}
{\Large \bf Some simple models for quark stars} \\
\vspace{1.5cm} {\bf S. D. Maharaj, J. M. Sunzu\footnote{Permanent address: School of Mathematical Sciences, University of Dodoma, Tanzania.} and S. Ray}\\
Astrophysics and Cosmology Research Unit,\\
School of Mathematics, Statistics and Computer Science,\\
University of KwaZulu-Natal,\\
Private Bag X54001,\\
Durban 4000,\\
South Africa.\\
\end{center}

\begin{abstract} 
We find two new classes of exact solutions for the Einstein-Maxwell equations. The solutions are obtained by considering charged anisotropic 
matter with a linear equation of state consistent with quark stars. 
The field equations are integrated by specifying forms for the measure of anisotropy 
and a gravitational potential which are physically reasonable. The solutions found generalize the Mark-Harko model and the Komathiraj-Maharaj model. 
A graphical analysis indicates that the matter variables are well behaved.

\end{abstract}

\section{Introduction}
The first study of quark stars was performed by Itoh \cite{Itoh} for static matter in equilibrium. The physical processes governing the behaviour of quark 
matter with ultrahigh densities is still under investigation. Of special interest is the equation of state for quark matter. The phenomenology of the MIT 
bag model indicates that a linear form for the equation of state is possible with a nonzero bag constant. This is shown in the works by 
Chodos \textit{et al} \cite{Chodos}, Farhi and Jaffe \cite{Farhi} and Witten \cite{Witten}. The review of Weber \cite{Weber} highlights models of compact 
astrophysical objects composed of strange quark stars. Some recent investigations for compact objects with a quark equation of state include the treatments of 
Kalam \textit{et al} \cite{Kalam2} and Mafa Takisa and Maharaj \cite{Mafa}. The effect of the electromagnetic field on quark star was studied by Mak and Harko 
\cite{Mak} in the presence of a conformal symmetry. Sharma and Maharaj \cite{Sharma} considered the role of anisotropy for a specified mass distribution. 
Charged anisotropy matter with a linear equation of state, extendible to the more general nonlinear case, was analysed by Varela 
\textit{et al} \cite{Varela}. Other papers containing interesting features relating to charge and anisotropy are given in the references \cite{Thirukkanesh}-\cite{Esculpi}.

Mak and Harko \cite{Mak} found strange quark stars with isotropic pressures in the presence of charge. Komathiraj and Maharaj \cite{Komathiraj} presented a 
method of solving the Einstein-Maxwell system to produce new models of charged quark stars. In the present work we show that the Komathiraj and Maharaj 
method allows us to integrate the Einstein-Maxwell equations with anisotropic pressures and charge. Therefore we are able to generate new quark stars which 
are charged and anisotropic. Two new classes of solutions to the field equations are obtained by specifying the measure of anisotropy. Earlier solutions 
are shown to be contained in our results. A notable feature of our models is that we get the anisotropy to vanish, for particular parameter values, 
and isotropic pressures are regained. In many previous investigations the anisotropy is always present which is not desirable. 
A physical analysis indicates that the gravitational potentials and the matter variables are well behaved, and 
we can generate masses consistent with observations.
\section{The model \label{one}}
We intend to model the stellar interior with quark matter in general relativity. The spacetime geometry is static and spherically symmetric. 
The interior spacetime is represented by the line element
\begin{equation}
 ds^{2}=-e^{2\nu(r)}dt^{2}+e^{2\lambda(r)}dr^{2}+r^{2}(d\theta^{2}+\sin^{2}\theta d\phi^{2}),
\label{line-element}
\end{equation}
where $\nu(r)$ and $\lambda(r)$ are arbitrary functions representing gravity. The exterior spacetime is given by the Reissner-Nordstrom line element
\begin{equation}
 ds^{2}=-\left(1-\frac{2M}{r}+\frac{Q^{2}}{r^{2}}\right)dt^{2}+ \left(1-\frac{2M}{r}+\frac{Q^{2}}{r^{2}}\right)^{-1}dr^{2}
+r^{2}(d\theta^{2}+\sin^{2}\theta d\phi^{2}),
\label{line-element-exterior}
\end{equation}
where $M$ and $Q$ are the total mass and charge of the star respectively.
The energy momentum tensor for anisotropic charged fluid matter is of the form 
\begin{equation}
 T_{ij}=\mbox{diag}\left(-\rho-\frac{1}{2}E^{2},p_{r}-\frac{1}{2}E^{2},p_{t}+\frac{1}{2}E^{2},p_{t}+\frac{1}{2}E^{2}\right).
\label{Energy-mom tensor}
\end{equation}
In the above $\rho$ is energy density, $p_{r}$ is the radial pressure, $p_{t}$ is the tangential pressure, and $E$ is the electric field intensity. 
These quantities are measured relative to a comoving unit timelike fluid four-velocity $u^{a}$.

Then the Einstein-Maxwell equations can be written as
\begin{subequations}
\label{Emf}
\begin{eqnarray}
 \dfrac{1}{r^{2}}\left( 1-e^{-2\lambda}\right)+\dfrac{2\lambda^\prime}{r}e^{-2\lambda}&=&\rho + \frac{1}{2}E^{2}, \\
\label{Emf1}
 -\dfrac{1}{r^{2}}\left( 1-e^{-2\lambda}\right)+\dfrac{2\nu^{\prime}}{r}e^{-2\lambda}&=&p_{r} - \frac{1}{2}E^{2},\\
\label{Emf2}
e^{-2\lambda}\left( \nu^{\prime\prime}+\nu^{\prime^{2}}-\nu^{\prime}\lambda^{\prime}+\dfrac{\nu^{\prime}}{r}-
\dfrac{\lambda^{\prime}}{r}\right)&=&p_{t}+ \frac{1}{2}E^{2},\\
\label{Emf3}
 \sigma&=&\frac{1}{r^{2}}e^{-\lambda}\left( r^{2}E\right)^{\prime},
\label{Emf4}
\end{eqnarray}
\end{subequations}
where primes denote differentiation with respect to radial coordinate $r$. The function $\sigma$ represents the proper charge density. 
We are using the units where the coupling constant and the speed of light are unity. For a quark star we assume a linear relationship 
between the radial pressure and the energy density
\begin{equation}
p_{r}=\frac{1}{3}\left(\rho-4B\right),
\label{eqnstate}
\end{equation}
where $B$ is the bag constant. To transform the field equations to a more convenient form we introduce new variables defined by
\begin{equation}
 x=Cr^{2},\;\;Z(x)=e^{-2\lambda(r)},\;\;A^{2}y^{2}(x)=e^{2\nu(r)},
\label{transformation}
\end{equation}
where $A$ and $C$ are arbitrary constants. Then the Einstein-Maxwell field equations (\ref{Emf}) for quark matter have the following form 
\begin{subequations}
\label{nnewemf}
\begin{eqnarray}
 \rho&=&3p_{r}+4B, \label{nnewemf1}\\
 \frac{p_{r}}{C}&=&Z\frac{\dot{y}}{y}-\frac{\dot{Z}}{2}-\frac{B}{C}, \label{nnewemf2}\\
  p_{t}&=&p_{r}+\Delta,\label{nnewtangetial}\\
 \Delta &=&\frac{4xCZ\ddot{y}}{y}+C\left(2x\dot{Z}+6Z\right)\frac{\dot{y}}{y}\nonumber \\
 & & +C\left(2\left(\dot{Z}+\frac{B}{C}\right)+\frac{Z-1}{x}\right),\label{nnewemf4}\\ 
\frac{E^{2}}{2C}&=&\frac{1-Z}{x}-3Z\frac{\dot{y}}{y}-\frac{\dot{Z}}{2}-\frac{B}{C},\label{nnewemf3}\\ 
\sigma&=& 2\sqrt{\frac{ZC}{x}}\left(x\dot{E}+E\right).\label{nnewemf5}
\end{eqnarray}
\end{subequations}
The quantity $\Delta= p_{t}-p_{r}$ is the measure of anisotropy. This system consists of eight variables 
$(\rho,\;p_{r},\;p_{t},\;E,\;Z,\;y,\;\sigma,\;\Delta)$ in six equations. It is apparent that if we specify two of these variables then the system may be integrated. 
The gravitational behavior of the anisotropic charged quark star is governed by 
the system (\ref{nnewemf}). For $\Delta=0$ we have the isotropic model that was described by Komathiraj and Maharaj \cite{Komathiraj}. For neutral fluids 
with isotropic pressures $(\Delta=0,\;E=0)$ there is no freedom in the system (\ref{nnewemf}) as the equation of state has been specified. For a charged 
fluid with anisotropic pressures $(\Delta \neq 0,\;E \neq 0)$, with the linear equation of state, there are two degrees of freedom because of 
the appearance of new matter quantities, the electric field and anisotropy. From a mathematical viewpoint any two of the eight variables may be chosen to 
integrate the system (\ref{nnewemf}); the choice should be carefully made on physical grounds so that a well behaved model results.

In order to find  exact solutions to this model we have to specify two quantities: we choose the potential $y$ and the quantity $\Delta$. 
We specify the metric function  
\begin{equation}
y=\left(a+x^{m}\right)^{n},
\label{Choice-y}
\end{equation}
where $a$, $m$ and $n$ are constants. A similar choice was made by Komathiraj and Maharaj \cite{Komathiraj}. The choice guarantees that 
the metric function $y$ is regular and well behaved within the interior. It remains finite at the centre of the star. Note that special cases of 
the potential $y$ corresponds to known quark models, e.g. when $m=\frac{1}{2}$, $n=1$ we regain the Mak and Harko \cite{Mak} quark star model and 
when $m=1$, $n=2$ we regain the Komathiraj and Maharaj \cite{Komathiraj} model for a quark star with isotropic pressures. We expect that the potential 
(\ref{Choice-y}) is therefore likely to produce new solutions, and more general models, to the Einstein-Maxwell system when charge and anisotropy are present. Also we specify the 
measure of anisotropy in the form
\begin{equation}
\Delta=A_{0}+A_{1}x+A_{2}x^{2}+A_{3}x^{3},
\label{choice-delta}
\end{equation}
where $A_{0},\;A_{1},\;A_{2},\text{and}\,A_{3\;}$ are arbitrary constants. This choice is physically reasonable and ensures that we regain isotropic 
pressures when $A_{0}=A_{1}=A_{2}=A_{3}=0$. Note that we have effectively taken three orders of a Taylor expansion for $\Delta$ in terms of the radial coordinate. 
This form of $\Delta$ enables us to integrate the Einstein-Maxwell system; higher order terms lead to expressions which are not integrable. An important 
point to note is that the form (\ref{choice-delta}) allows us to regain isotropic pressures by setting parameters to vanish. In most other treatments 
involving anisotropic stellar configurations this is not the case as indicated in the work of Dev and Gleiser \cite{Dev2}, Esculpi and Aloma \cite{Esculpi}, 
Harko and Mak \cite{Harko}, and Mak and Harko \cite{Mak2}. The recent strange quark models of Kalam \textit{et al} \cite{Kalam2} and Paul 
\textit{et al} \cite{Paul} also have a nonzero anisotropy throughout the star. In our model the choice (\ref{choice-delta}) enables us to regain isotropic pressures. 
From (\ref{Choice-y}) and (\ref{nnewemf4}), and simplifying using partial fractions, we obtain the 
differential equation
\begin{equation}
\begin{split}
\dot{Z}+
\left( \frac{1}{2x}+\dfrac{2m(n-1)x^{m-1}}{a+x^{m}}+\dfrac{m\left( 4(1+mn)-3n\right) x^{m-1}}{2\left( a+(1+mn)x^{m}\right)}\right)Z
\\=
\dfrac{\left(  1-\frac{2xB}{C}+\frac{(A_{0}+A_{1}x+A_{2}x^{2}+A_{3}x^{3})x}{C}\right) \left( a+x^{m}\right) }{2x \left( a+(1+mn)x^{m}\right)}.
\end{split}
\label{ode4}
\end{equation}
Once (\ref{ode4}) is integrated we can directly find the remaining quantities $\rho$, $p_{r}$, $p_{t}$, $E^{2}$ and $\sigma$ from the 
system (\ref{nnewemf}). In order to find an exact solution to (\ref{ode4}) we need to specify values for the constants $m$ and $n$.

\section{Generalized Komathiraj-Maharaj model \label{four}}
We can find an exact solution to (\ref{ode4}) when $m=\frac{1}{2}$ and $n=1$. In this case line element becomes
\begin{eqnarray}
ds^{2}&=&-A^{2}\left(a+\sqrt{x}\right)^{2}dt^{2}+
\left(\dfrac{3 \left( 2a+3\sqrt{x} \right)}{ 3\left( 2a+\sqrt{x}\right)-\frac{Bx}{C} \left( 4a+3\sqrt{x} \right)+\frac{3F(x)}{C}}\right)dr^{2}\nonumber\\
& & +r^{2}(d\theta^{2}+\sin^{2}\theta d\phi^{2}).
\label{mk-line element}
\end{eqnarray}
The matter variables are qiven by 
\begin{subequations}
\label{exactsoln-mak-harko}
\begin{eqnarray}
\rho&=&\dfrac{3C\left(6a^{2}+10a\sqrt{x}+3x\right)}{2\sqrt{x}(a+\sqrt{x})(2a+3\sqrt{x})^{2}}\nonumber
+\dfrac{B\left( 16a^{3}+47a^{2}\sqrt{x}+48ax+18x^{\frac{3}{2}}\right) }{2(a+\sqrt{x})(2a+3\sqrt{x})^{2}}\\
& &-3G(x)\left( \dfrac{1}{2\left( a+ \sqrt{x}\right)\left( 2a+ 3\sqrt{x}\right)^{2}}\right),\label{mak-energydensity}\\
 p_{r}&=&\dfrac{C\left(6a^{2}+10a\sqrt{x}+3x\right)}{2\sqrt{x}(a+\sqrt{x})(2a+3\sqrt{x})^{2}}\nonumber
-\dfrac{B\left( \frac{16}{3}a^{3}+27a^{2}\sqrt{x}+40ax+18x^{\frac{3}{2}}\right) }{2(a+\sqrt{x})(2a+3\sqrt{x})^{2}}\\ 
& &-G(x)\left( \dfrac{1}{2\left( a+ \sqrt{x}\right)\left( 2a+ 3\sqrt{x}\right)^{2}}\right),\label{mak-radialpressure}\\
 p_{t}&=&\dfrac{C\left(6a^{2}+10a\sqrt{x}+3x\right)}{2\sqrt{x}(a+\sqrt{x})(2a+3\sqrt{x})^{2}} \nonumber
-\dfrac{B\left( \frac{16}{3}a^{3}+27a^{2}\sqrt{x}+40ax+18x^{\frac{3}{2}}\right) }{2(a+\sqrt{x})(2a+3\sqrt{x})^{2}}\\
& & + H(x) \left( \dfrac{1}{2\left( a+ \sqrt{x}\right)\left( 2a+ 3\sqrt{x}\right)^{2}}\right),\label{mak-tangentialpressure}\\
\Delta &=& A_{0}+A_{1}x+A_{2}x^{2}+A_{3}x^{3},\\
 E^{2}&=&\left[ C\left( -2a^{2}-2a\sqrt{x}+3x\right)+Bx\left( a^{2}+2a\sqrt{x}\right) -J(x)\right]\nonumber \\  
& & \times \left( \frac{1}{\sqrt{x}( a+\sqrt{x})(2a+3\sqrt{x})^{2}}\right). \label{mak-electricfield}
\end{eqnarray}
\end{subequations}
To simplify the relevant expressions we have set 
\begin{eqnarray*}
F(x)&=& A_{0}\left( \frac{2}{3} ax+\frac{1}{2}x^{\frac{3}{2}}\right)+A_{1}\left( \frac{2}{5}ax^{2}+\frac{1}{3}x^{\frac{5}{2}}\right)\\
 & & + A_{2}\left(\frac{2}{7}ax^{3}+\frac{1}{4}x^{\frac{7}{2}}\right)
  +A_{3}\left( \frac{2}{9}ax^{4}+\frac{1}{5}x^{\frac{9}{2}}\right),\\
G(x) &=& A_{0}\left(  \frac{4}{3}a^{3}+\frac{5}{2}a^{2}\sqrt{x}+ax \right)\\
& &+A_{1}\left( \frac{8}{5}a^{3}x+\frac{64}{15}a^{2}x^{\frac{3}{2}}+\frac{18}{5}ax^{2}+x^{\frac{5}{2}}\right)\\
& &+A_{2}\left( \frac{12}{7}a^{3}x^{2}+\frac{141}{28}a^{2}x^{\frac{5}{2}}+\frac{67}{14}ax^{3}+\frac{3}{2}x^{\frac{7}{2}}\right)\\
& &+A_{3}\left(  \frac{16}{9}a^{3}x^{3}+\frac{82}{15}a^{2}x^{\frac{7}{2}}+\frac{82}{15}ax^{4}+\frac{9}{5}x^{\frac{9}{2}}\right),\\
H(x) &=& A_{0}\left(  \frac{20}{3}a^{3}+\frac{59}{2}a^{2}\sqrt{x}+41ax+18x^{\frac{3}{2}} \right)\\
& &+A_{1}\left( \frac{32}{5}a^{3}x+\frac{416}{15}a^{2}x^{\frac{3}{2}}+\frac{192}{5}ax^{2}+17x^{\frac{5}{2}}\right)\\
& & +A_{2}\left( \frac{44}{7}a^{3}x^{2}+\frac{755}{28}a^{2}x^{\frac{5}{2}}+\frac{521}{14}ax^{3}+\frac{33}{2}x^{\frac{7}{2}}\right)\\
& &+A_{3}\left(  \frac{56}{9}a^{3}x^{3}+\frac{398}{15}a^{2}x^{\frac{7}{2}}+\frac{548}{15}ax^{4}+\frac{81}{5}x^{\frac{9}{2}}\right),\\
J(x)&=&A_{0}\left(  4a^{3}\sqrt{x} +\frac{33}{2}a^{2}x+22ax^{\frac{3}{2}}+9x^{2} \right)\\
& &+A_{1}\left( \frac{16}{5}a^{3}x^{\frac{3}{2}}+\frac{64}{5}a^{2}x^{2}+\frac{84}{5}ax^{\frac{5}{2}}+7x^{3}\right)\\
& &+A_{2}\left( \frac{20}{7}a^{3}x^{\frac{5}{2}}+\frac{313}{28}a^{2}x^{3}+\frac{101}{7}ax^{\frac{7}{2}}+6x^{4}\right)\\
& &+A_{3}\left(  \frac{8}{3}a^{3}x^{\frac{7}{2}}+\frac{154}{15}a^{2}x^{4}+\frac{196}{15}ax^{\frac{9}{2}}+\frac{27}{5}x^{5}\right).
\end{eqnarray*}
The exact solution (\ref{mk-line element}) and (\ref{exactsoln-mak-harko}) is a new model for a charged anisotropic quark star. 
If we set $A_{0}=A_{1}=A_{2}=A_{3}=0$, then we regain the first Komathiraj and Maharaj \cite{Komathiraj}
line element 
\begin{eqnarray}
ds^{2}&=&-A^{2}\left(a+\sqrt{x}\right)^{2}dt^{2}+
\left(\dfrac{3 \left( 2a+3\sqrt{x} \right)}{ 3\left( 2a+\sqrt{x}\right)-\frac{Bx}{C} \left( 4a+3\sqrt{x} \right)}\right)dr^{2}\nonumber\\
& & +r^{2}(d\theta^{2}+\sin^{2}\theta d\phi^{2}),
\label{mk-line element2}
\end{eqnarray}
with isotropic pressures and with equation of state $p=\frac{1}{3}\left(\rho-4B\right)$. We observe that when we set $G(x)=0$, $H(x)=0$ and $J(x)=0$ 
in ({\ref{exactsoln-mak-harko}}) we obtain expressions 
for the energy density $\rho$, the pressure $p$ and electric field $E^{2}$ which are identical to those in the Komathiraj and Maharaj \cite{Komathiraj} 
model. Furthermore if we let $a=0$ in (\ref{mk-line element2}) we obtain 
Mak and Harko \cite{Mak} line element 
\begin{equation}
ds^{2}=-A^{2}Cr^{2}dt^{2}+\left(\frac{3}{1-Br^{2}}\right) dr^{2}+r^{2}(d\theta^{2}+\sin^{2}\theta d\phi^{2}),
\label{mk-line element3}
\end{equation}
with the matter variables
\begin{equation*}
\rho=\frac{1}{2r^{2}}+B,\;\;
 p=\frac{1}{6r^{2}}-B,\;\;E^{2}=\frac{1}{3r^{2}}.
\label{mk-particular}
\end{equation*}
On setting $B=0$ we regain the Misner and Zapolsky \cite{Misner} particular solution with the equation of state $p=\frac{1}{3}\rho$. 
Note that the class of solutions found in this section contains a singularity in the electric field at the centre. This feature is also present 
in the Mak and Harko \cite{Mak} model for quark star. However the total charge and mass remains finite which is a good  feature of this class of models.

\section{Nonsingular quark model \label{five}}
We can find another exact solution of (\ref{ode4}) by choosing $m=1$ and $n=2$. For this choice the line element becomes 
\begin{equation}
\begin{aligned}
ds^{2}=&-A^{2}\left(a+x\right)^{4}dt^{2}+r^{2}(d\theta^{2}+\sin^{2}\theta d\phi^{2})\\
& +\dfrac{{315( a+x)^{2}(a+3x)}dr^{2}}{9\left( 35a^{3}+35a^{2}x+21ax^{2} + 5x^{3} \right)
-\dfrac{2Bx}{C}\left( 105a^{3}+189a^{2}x+135ax^{2} + 35x^{3} \right)+\frac{315 L(x)}{C}}.
\label{nonsingular-line element1}
\end{aligned}
\end{equation}
The matter variables are 
\begin{subequations}
\begin{eqnarray}
\rho&=&\dfrac{3C\left( 140a^{4}+434a^{3}x+318a^{2}x^{2} + 150ax^{3}+30x^{4}\right)}{35( a+x)^{3}(a+3x)^{2}}\nonumber\\
& &+\dfrac{3\Psi(x)+B\left( 210a^{5}+798a^{4}x+1476a^{3}x^{2} + 2540a^{2}x^{3}+2090 ax^{4} +630 x^{5} \right) }{105( a+x)^{3}(a+3x)^{2}},\label{nonsingular-energydensity}\\
p_{r}&=&\dfrac{C\left( 140a^{4}+434a^{3}x+318a^{2}x^{2} + 150ax^{3}+30x^{4}\right)}{35( a+x)^{3}(a+3x)^{2}}\nonumber\\
& &+\dfrac{\Psi(x)-B\left( 70a^{5}+994a^{4}x+3708a^{3}x^{2} + \frac{16780}{3}a^{2}x^{3}+\frac{11770}{3}ax^{4} +1050x^{5} \right) }{105( a+x)^{3}(a+3x)^{2}},\label{nonsingular-radialp}\\
 p_{t}&=&\dfrac{C\left( 140a^{4}+434a^{3}x+318a^{2}x^{2} + 150ax^{3}+30x^{4}\right)}{35( a+x)^{3}(a+3x)^{2}}\nonumber\\
& & +\dfrac{\Omega(x)-B\left( 70a^{5}+994a^{4}x+3708a^{3}x^{2} + \frac{16780}{3}a^{2}x^{3}+\frac{11770}{3}ax^{4} +1050x^{5} \right) }{105( a+x)^{3}(a+3x)^{2}},\label{nonsingular-tangentialp}\\
\Delta&=&A_{0}+A_{1}x+A_{2}x^{2}+A_{3}x^{3},\\
 E^{2}&=&\dfrac{C\left(1764a^{3}x+13068a^{2}x^{2}+12204ax^{3} + 3780x^{4}\right)-\Lambda(x)}{315( a+x)^{3}(a+3x)^{2}}\nonumber\\
& &-\dfrac{B\left(168a^{4}x+1296a^{3}x^{2}+6528a^{2}x^{3}+7280ax^{4}+2520x^{5}\right) }{315( a+x)^{3}(a+3x)^{2}}.
\label{nonsingular-electricf}
\end{eqnarray}
\label{nonsingular-exact}
\end{subequations}
For simplicity we have set
\begin{eqnarray*}
L(x)&=&A_{0}\left( \frac{1}{3}a^{3}x+\frac{3}{5}a^{2}x^{2}+\frac{3}{7}ax^{3}+\frac{1}{9}x^{4} \right)\\
 & &+A_{1}\left( \frac{1}{5}a^{3}x^{2}+\frac{3}{7}a^{2}x^{3}+\frac{1}{3}ax^{4}+\frac{1}{11}x^{5}\right)\\
 & &+A_{2}\left( \frac{1}{7}a^{3}x^{3}+\frac{1}{3}a^{2}x^{4}+\frac{3}{11}ax^{5}+\frac{1}{13}x^{6}\right)\\
& &+A_{3}\left(  \frac{1}{9}a^{3}x^{4}+\frac{3}{11}a^{2}x^{5}+\frac{3}{13}ax^{6}+\frac{1}{15}x^{7}\right),\\
\Psi(x) &=& A_{0}\left( -\frac{35}{2}a^{5}+\frac{49}{2}a^{4}x+279a^{3}x^{2}+\frac{1145}{3}a^{2}x^{3}+\frac{1375}{6}ax^{4}+\frac{105}{2}x^{5}\right)\\
& & +A_{1}x\left(-21a^{5}-57a^{4}x+20a^{3}x^{2}+\frac{1360}{11}a^{2}x^{3}+105ax^{4} +\frac{315}{11}x^{5} \right)\\
& & +A_{2}x^{2}\left(-\frac{45}{2}a^{5}-\frac{185}{2}a^{4}x-\frac{1145}{11}a^{3}x^{2}-\frac{315}{13}a^{2}x^{3}+\frac{7245}{286}ax^{4}+ \frac{315}{26}x^{5}\right)\\
& & -A_{3}x^{3}\left(\frac{70}{3}a^{5}+\frac{3710}{33}a^{4}x+\frac{2310}{13}a^{3}x^{2}+\frac{17206}{143}a^{2}x^{3}+\frac{392}{13}ax^{4} \right),\\
\Omega(x) &=& A_{0}\left( \frac{175}{2}a^{5}+\frac{1939}{2}a^{4}x+3429a^{3}x^{2}+\frac{15635}{3}a^{2}x^{3}+\frac{22165}{6}ax^{4}+\frac{1995}{2}x^{5}\right)\\
& & +A_{1}x\left(84a^{5}+888a^{4}x+3170a^{3}x^{2}+\frac{54490}{11}a^{2}x^{3}+3570ax^{4} +\frac{10710}{11}x^{5} \right)\\
& & +A_{2}x^{2}\left(\frac{165}{2}a^{5}+\frac{1705}{2}a^{4}x+\frac{33505}{11}a^{3}x^{2}+\frac{62474}{13}a^{2}x^{3}+\frac{998235}{286}ax^{4}+\frac{24885}{26}x^{5}\right)\\
& & +A_{3}x^{3}\left(\frac{245}{3}a^{5}+\frac{27475}{33}a^{4}x+\frac{38640}{13}a^{3}x^{2}+\frac{673484}{143}a^{2}x^{3}+\frac{44653}{13}ax^{4}+945x^{5}\right),\\
\Lambda(x)&=&A_{0}\left( 315a^{5}+2751a^{4}x+8802a^{3}x^{2}+11226a^{2}x^{3}+6755ax^{4}+1575x^{5}\right)\\
& &+A_{1}x\left(252a^{5}+2124a^{4}x+6732a^{3}x^{2}+\frac{100380}{11}a^{2}x^{3}+\frac{63000}{11}ax^{4} +\frac{15120}{11}x^{5} \right)\\
& &+A_{2}x^{2}\left(225a^{5}+1845a^{4}x+\frac{63210}{11}a^{3}x^{2}+\frac{1133370}{143}a^{2}x^{3}+\frac{55755}{11}ax^{4}+\frac{16065}{13}x^{5}\right)\\
& &+A_{3}x^{3}\left(210a^{5}+\frac{18550}{11}a^{4}x+\frac{738360}{143}a^{3}x^{2}+\frac{78624}{11}a^{2}x^{3}+\frac{59934}{13}ax^{4}+1134x^{5} \right).
\end{eqnarray*}
The mass function giving the total mass within a sphere of radius $x$ is given by 
\begin{equation}
\begin{split}
\begin{aligned}
m(x)=& \left(\left(\frac{1268}{96525}a-\frac{1}{30}x^{2}-\frac{14}{585}ax\right) A_{3}-\left(\frac{4}{91}x+\frac{74}{2145}a\right) A_{2}
-\frac{7}{110}A_{1}\right)\frac{x^{\frac{5}{2}}}{C^\frac{3}{2}} \\
&-\left(\frac{1}{9}B+\frac{1}{9}A_{0}+\frac{2}{33}aA_{1}-\frac{10}{429}a^{2}A_{2}+ \frac{14}{1287}a^{3}A_{3}\right)\left(\frac{x}{C}\right)^{\frac{3}{2}}\\
&-\sqrt{\frac{a}{C^{3}}}\left( \frac{62}{105}aB+\frac{93}{35}C-\frac{31}{105}aA_{0}+\frac{31}{385}a^{2}A_{1}-\frac{31}{1001}a^{3}A_{2}
+\frac{31}{2145}a^{4}A_{3}\right)\arctan \sqrt{\frac{x}{a}}\\ 
&+\frac{\sqrt{3a}}{3C^\frac{3}{2}}\left( \frac{188}{315}aB+\frac{129}{35}C-\frac{94}{315}aA_{0}+\frac{59}{1155}a^{2}A_{1}-\frac{100}{9009}a^{3}A_{2}
+\frac{157}{57915}a^{4}A_{3}\right)\arctan\sqrt{\frac{3x}{a}}\\
&+\left(\frac{76}{189}aB+\frac{8}{9}C-\frac{38}{189}aA_{0}+\frac{52}{693}a^{2}A_{1}-\frac{934}{27027}a^{3}A_{2}
+\frac{3088}{173745}a^{4}A_{3}\right)\sqrt{\frac{x}{C^{3}}}\\
&-\left(\frac{6}{35}a^{2}B+\frac{27}{35}aC-\frac{3}{35}a^{2}A_{0}+\frac{9}{385}a^{3}A_{1}-\frac{9}{1001}a^{4}A_{2}
+\frac{3}{715}a^{5}A_{3}\right)\frac{\sqrt{x}}{(a+x)C^\frac{3}{2}}\\
&-\left(\frac{4}{105}a^{3}B+\frac{6}{35}a^{2}C-\frac{2}{105}a^{3}A_{0}+\frac{2}{385}a^{4}A_{1}-\frac{2}{1001}a^{5}A_{2}+
\frac{2}{2145}a^{6}A_{3}\right)\frac{\sqrt{x}}{(a+x)^{2}C^{\frac{3}{2}}}\\
&-\left(\frac{188}{945}a^{2}B+\frac{43}{35}aC-\frac{94}{945}a^{2}A_{0}+\frac{59}{3465}a^{3}A_{2}-\frac{100}{27027}a^{4}A_{2}+
\frac{157}{173745}a^{5}A_{3}\right)\frac{\sqrt{x}}{(a+3x)C^{\frac{3}{2}}}
\end{aligned}
\end{split}
\end{equation}

The exact solution (\ref{nonsingular-line element1}) and (\ref{nonsingular-exact}) is a new model for the Einstein-Maxwell system with charge and 
anisotropy. If we set $A_{0}=A_{1}=A_{2}=A_{3}=0$, then we regain the second Komathiraj and Maharaj \cite{Komathiraj} line element 
\begin{equation}
\begin{aligned}
ds^{2}=&-A^{2}\left(a+x\right)^{4}dt^{2}+r^{2}(d\theta^{2}+\sin^{2}\theta d\phi^{2})\\
& +\dfrac{{315( a+x)^{2}(a+3x)}dr^{2}}{9\left( 35a^{3}+35a^{2}x+21ax^{2} + 5x^{3} \right)
-\dfrac{2Bx}{C}\left( 105a^{3}+189a^{2}x+135ax^{2} + 35x^{3} \right)}
\label{nonsingular-line element2}
\end{aligned}
\end{equation}
with isotropic pressures. We observe that when we set $\Psi(x)=0$, $\Omega(x)=0$ and $\Lambda(x)=0$ in ({\ref{nonsingular-exact}}) we obtain expressions 
for the energy density $\rho$, the pressure $p$ and electric field $E^{2}$ corresponding to the nonsingular Komathiraj and Maharaj \cite{Komathiraj} 
model. The matter variables, including the electric field, remain finite at the centre so that the model is nonsingular. 
Consequently the class of solutions found in this section are good candidates to produce charged anisotropic stars with physically reasonable interior 
distributions.

\section{Discussion \label{six}}
\begin{figure}[h!]
\psfrag{radial distance}{radial distance $r$}
\psfrag{potential}{potential $e^{2\nu}$}
 \centering
 \includegraphics[width=10cm,height=8cm]{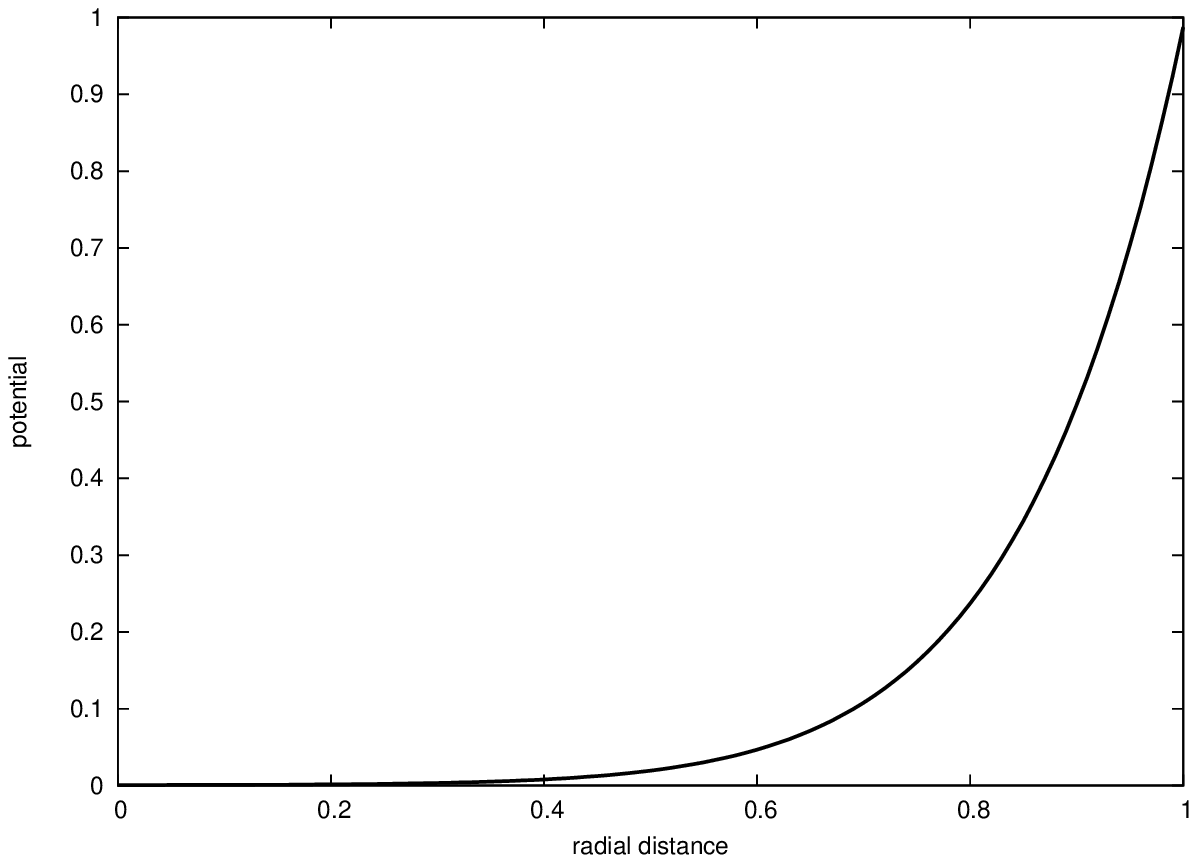}
 \caption{Potential $e^{2\nu}$ against the radial distance $r$}
 \label{one}
\psfrag{radial distance}{ radial distance $r$}
\psfrag{potential}{potential $e^{2\lambda}$}
 \centering
 \includegraphics[width=10cm,height=8cm]{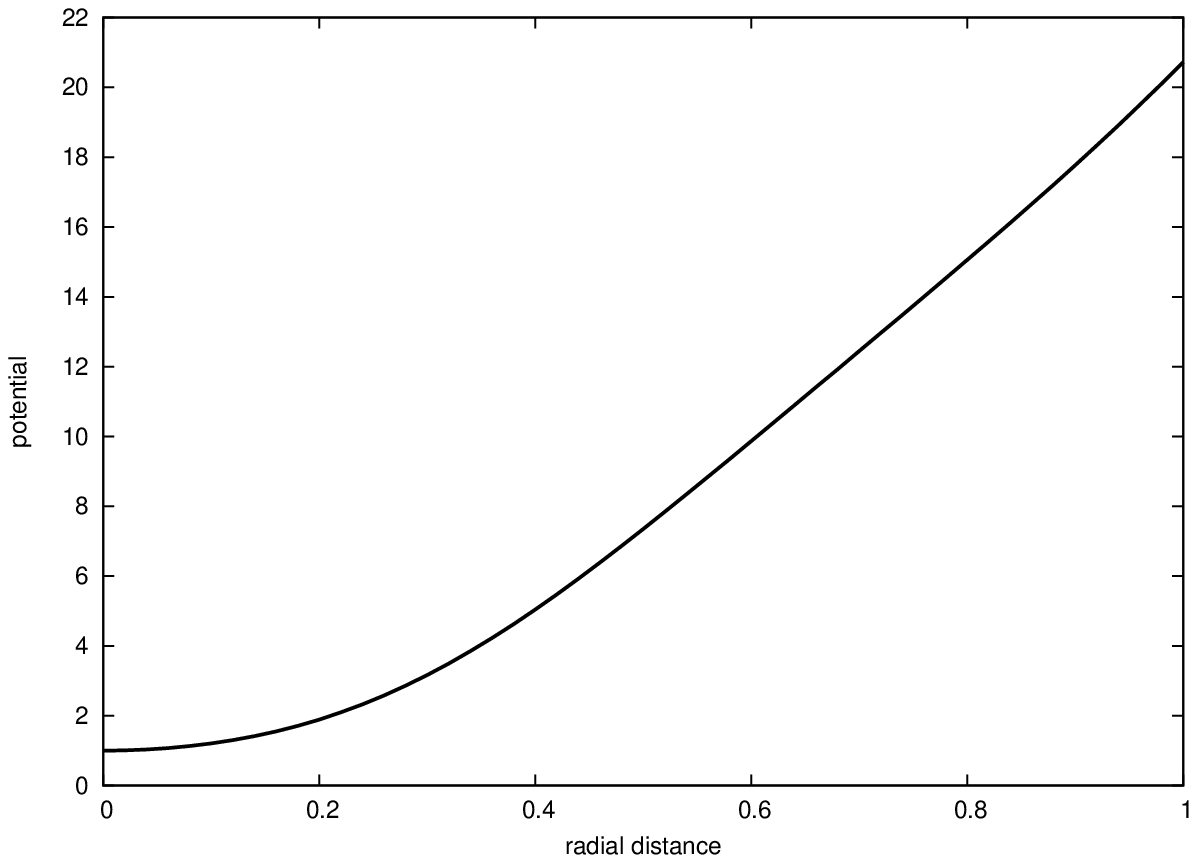}
 \caption{Potential $e^{2\lambda}$ against the radial distance $r$}
 \label{two}
\end{figure}

\begin{figure}[h!]
\psfrag{radial distance}{ radial distance $r$}
\psfrag{energy density}{energy density $\rho$}
 \centering
 \includegraphics[width=10cm,height=8cm]{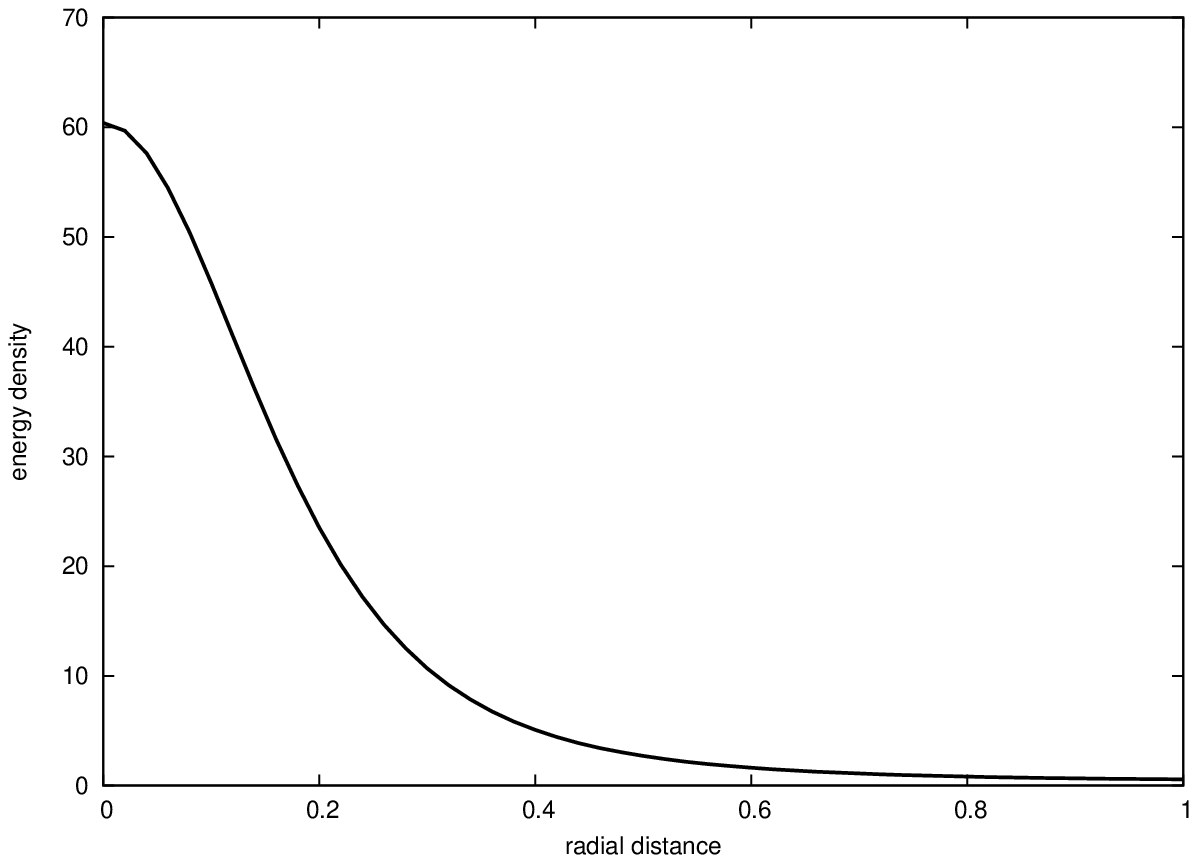}
 \caption{Energy density $\rho$ against the radial distance $r$}
 \label{three}
\psfrag{radial distance}{ radial distance $r$}
\psfrag{radial pressure}{radial pressure $p_{r}$}
 \centering
 \includegraphics[width=10cm,height=8cm]{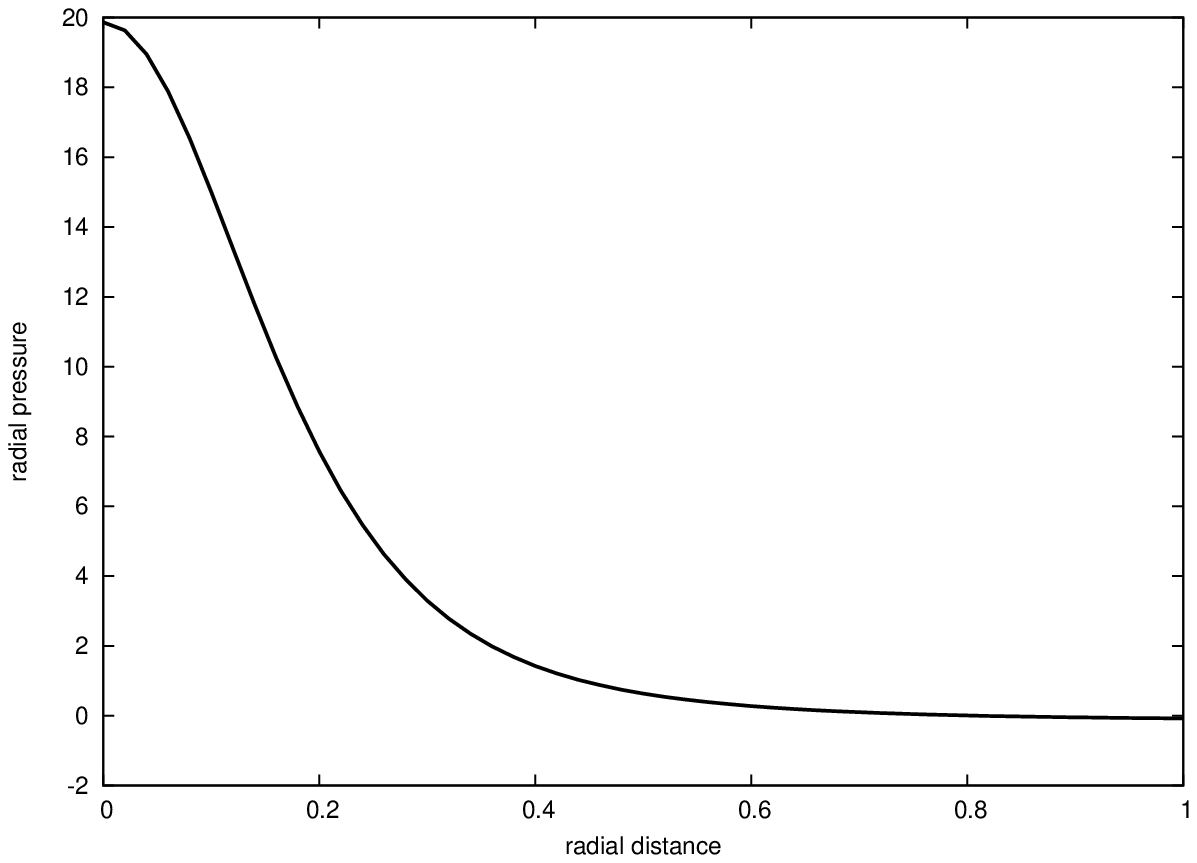}
 \caption{Radial pressure $p_{r}$ against radial distance $r$}
 \label{four}
\end{figure}

\begin{figure}[h!]
\psfrag{radial distance} {radial distance $r$}
\psfrag{electric field}{electric field $E^{2}$}
 \centering
 \includegraphics[width=10cm,height=8cm]{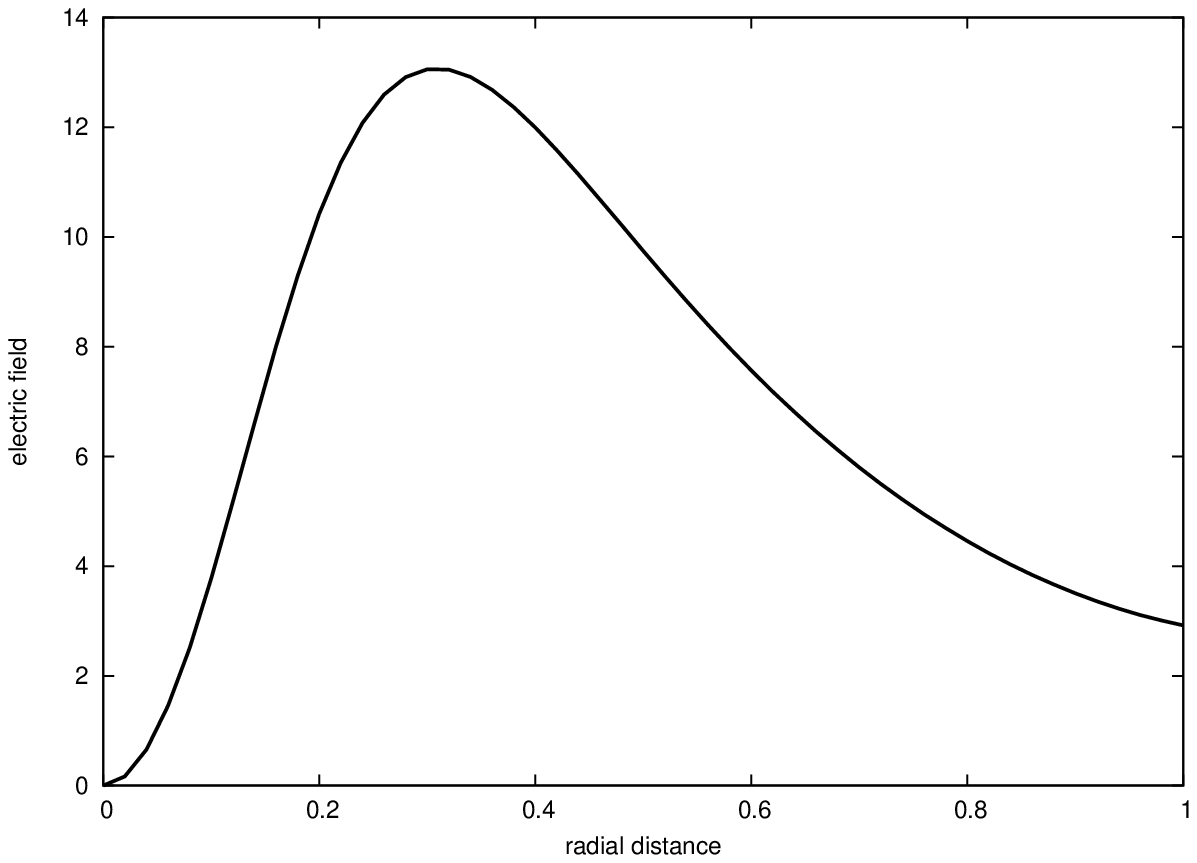}
 \caption{Electric field $E^{2}$ against radial distance $r$}
 \label{seven}
\psfrag{radial distance} {radial distance $r$}
\psfrag{mass}{mass $m$}
 \centering
 \includegraphics[width=10cm,height=8cm]{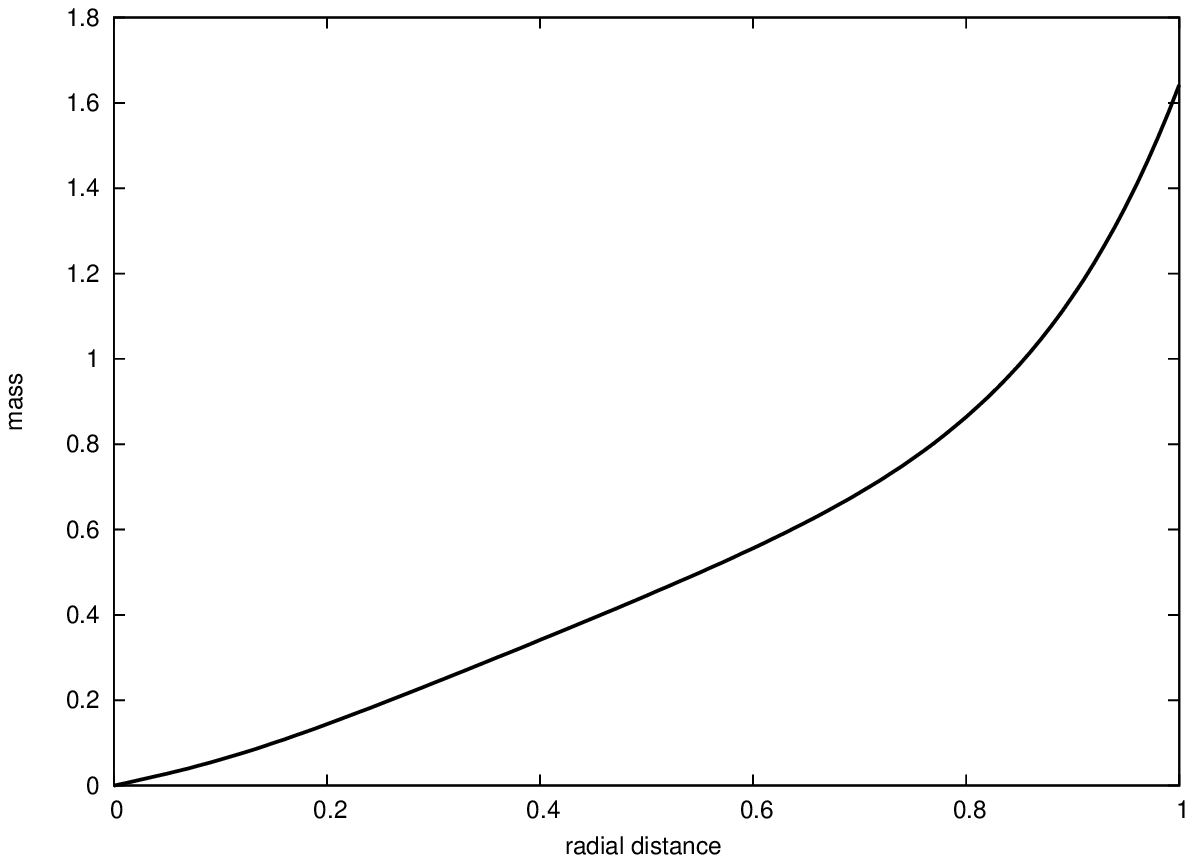}
 \caption{Mass $m$ against radial distance $r$}
 \label{eight}
\end{figure}

In this section we indicate that the exact solution of the field equations of the previous section are well behaved. 
To do this we generate graphical plots for the gravitational potentials, matter variables and the electric field. The Python programming language 
was used to generate plots for the particular choices $a=0.2$, $A=0.69$, $B=0.198$, $C=1$, $A_{0}=0.0$, $A_{1}=0.6$, $A_{2}=0.15$, and $A_{3}=-0.7$. 
The graphical plots generated are for the potential $e^{2\nu}$ (Fig. \ref{one}), potential $e^{2\lambda}$ (Fig. \ref{two}), energy density $\rho$ 
(Fig. \ref{three}), radial pressure $p{_r}$ (Fig. \ref{four}), tangential pressure $p_{t}$ (Fig. \ref{five}), measure of anisotropy 
$\Delta$ (Fig. \ref{six}), the electric field $E^{2}$ (Fig. \ref{seven}) and the mass $m$ (Fig. \ref{eight}). All figures are plotted against 
the radial coordinate $r$. These quantities are regular and well behaved in the stellar interior. The energy density, the radial pressure and the tangential 
pressure are decreasing functions as we approach the boundary from the centre. In general the measure of anisotropy $\Delta$ is finite and continuous. 
We observe that $\Delta$ increases from the centre until it attains a maximum value and decreases sharply towards the surface of the star. 
This profile is similar to that obtained by Sharma and Maharaj \cite{Sharma} and Mafa Takisa and Maharaj \cite{Mafa}. The electric field $E^{2}$ 
is finite and regular at the centre. It increases from the centre and then decreases after reaching a maximum value. We observe in Fig. \ref{eight} 
that the mass increases with radial distance monotonically. 

Finally we note for the values $a=0.0278$, $B=0.0064$, $C=0.0005$, $A_{0}=0.0000$, $A_{1}=0.0107$, $A_{2}=0.0134$, 
and $A_{3}=0.0107$ we can generate a quark star with radius $R=9.46$km and mass $M=2.86M_{\odot}$. These figures correspond 
to a distribution with a linear quark equation of state. They are consistent with the values found by Mak and Harko \cite{Mak}. 
Other values of the parameters produce radii and masses consistent with previous investigations. A detailed analysis of the physical 
features of the models found here will be undertaken in future work. 

\newpage

\newpage

\newpage

\begin{figure}[h!]
\psfrag{radial distance}{radial distance $r$}
\psfrag{tangential pressure}{tangential pressure $p_{t}$}
 \centering
 \includegraphics[width=10cm,height=8cm]{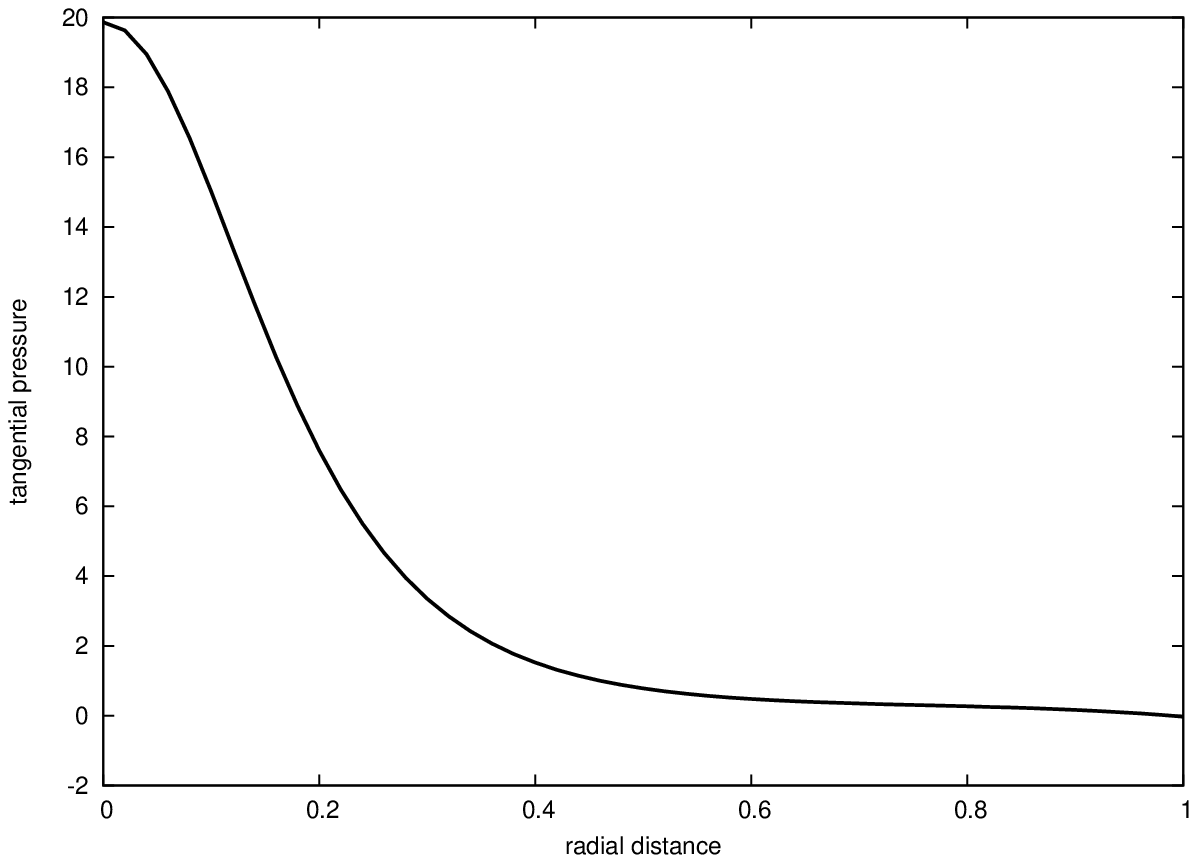}
 \caption{Tangential pressure $p_{t}$ against radial distance $r$}
 \label{five}
\psfrag{radial distance}{radial distance $r$}
\psfrag{anisotropy}{anisotropy $\Delta $}
 \centering
 \includegraphics[width=10cm,height=8cm]{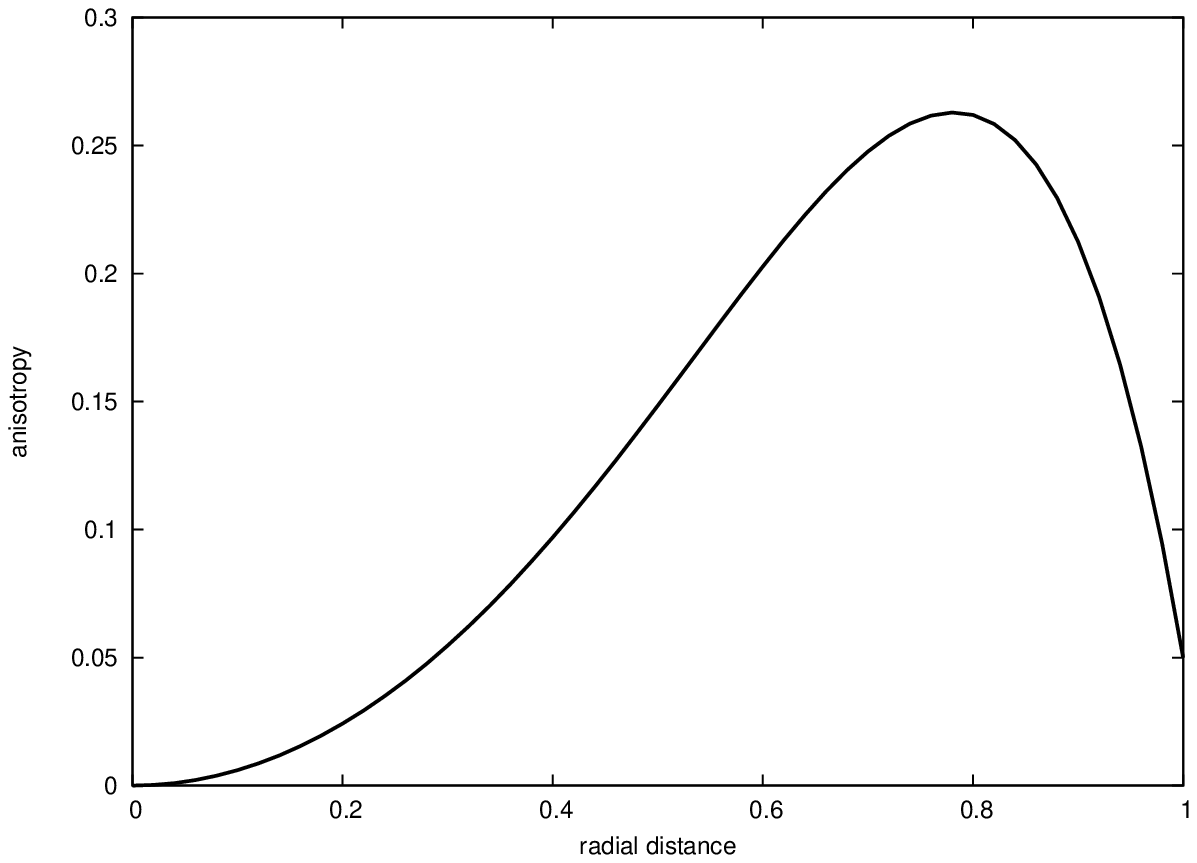}
 \caption{Measure of anisotropy $\Delta$ against radial distance $r$}
 \label{six}
\end{figure}

\newpage

\newpage

\begin{center}
\textbf{Acknowledgements }
\end{center}
We are grateful to the National Research Foundation and the University of KwaZulu-Natal for financial support. 
SDM acknowledges that this work is based upon research supported by the South African Research Chair Initiative of the 
Department of Science and Technology. JMS extends his appreciation to the University of Dodoma in Tanzania for study leave.

\end{document}